\documentclass[12pt]{article}

\usepackage{scicite}

\usepackage{times} 
\usepackage{graphicx}
\usepackage{amsmath}
\usepackage{color}

\newenvironment{sciabstract}{%
\begin{quote} \bf}
{\end{quote}}

\newcounter{lastnote}
\newenvironment{scilastnote}{%
\setcounter{lastnote}{\value{enumiv}}%
\addtocounter{lastnote}{+1}%
\begin{list}%
{\arabic{lastnote}.}
{\setlength{\leftmargin}{.22in}}
{\setlength{\labelsep}{.5em}}}
{\end{list}}

\title{The Sense of Place: \\
Grid Cells in the Brain and the Transcendental Number {\it e}}

\author
{Xue-Xin Wei,$^{1}$ Jason Prentice,$^{2,5}$ Vijay Balasubramanian$^{2,3,4\ast}$\\
\\
\normalsize{$^{1}$ Department of Psychology, University of Pennsylvania, Philadelphia, PA 19104, USA}\\
\normalsize{$^{2}$ Department of Physics, University of Pennsylvania, Philadelphia, PA 19104, USA}\\
\normalsize{$^{3}$ Department of Neuroscience, University of Pennsylvania, Philadelphia, PA 19104, USA}\\
\normalsize{$^{4}$ Laboratoire de Physique Th\'{e}orique, \'{E}cole Normale Sup\'{e}rieure, 75005 Paris, France}\\
\normalsize{$^{5}$ Princeton Neuroscience Institute, Princeton University, Princeton, NJ 08544, USA}\\
\\
\normalsize{$^\ast$Address correspondence to: vijay@physics.upenn.edu.}
}

\date{}

\begin{document} 


%\baselineskip24pt


\maketitle


\begin{sciabstract}

Grid cells in the brain respond when an animal occupies a periodic lattice of  ``grid fields" during spatial navigation.  The grid scale varies along the dorso-ventral axis of the entorhinal cortex.  We propose that the grid system minimizes the number of neurons required to encode location with a given resolution.  We derive several predictions that match recent experiments: (i) grid scales follow a geometric progression, (ii) the ratio between adjacent grid scales is $\sqrt{e}$ for idealized neurons, and robustly lies in the range 1.4-1.7 for realistic neurons, (iii) the scale ratio varies modestly within and between animals,  (iv) the ratio between grid scale and individual grid field widths at that scale also lies in this range, (v) grid fields lie on a triangular lattice. The theory also predicts the optimal grids in one and three dimensions, and the total number of discrete scales.   

\end{sciabstract}

% In setting up this template for *Science* papers, we've used both
% the \section* command and the \paragraph* command for topical
% divisions.  Which you use will of course depend on the type of paper
% you're writing.  Review Articles tend to have displayed headings, for
% which \section* is more appropriate; Research Articles, when they have
% formal topical divisions at all, tend to signal them with bold text
% that runs into the paragraph, for which \paragraph* is the right
% choice.  Either way, use the asterisk (*) modifier, as shown, to

%\section*{aa}

\section*{Introduction}

How does the brain represent space?   Tolman\cite{Tolman1948} suggested that the brain must have an explicit neural representation of physical space, a {\it cognitive map}, that supports higher brain functions such as navigation and path planning.    The discovery of place cells in the rat hippocampus \cite{OÕKeefe1976,OÕKeefe1978} suggested one potential locus for this map.  Place cells have spatially localized firing fields which reorganize dramatically when the environment changes \cite{Leutgeb2005}.  Another potential locus for the cognitive map of space has been uncovered in the main input to hippocampus, a structure known as the medial entorhinal cortex (MEC) \cite{Fyhn2004,Hafting2005}.   When rats freely explore a two dimensional open environment, individual ``grid cells'' in the MEC display spatial firing fields that form a periodic triangular grid which tiles space (Fig.~1A).     It is believed that  grid fields provide relatively rigid coordinates on space based partly on self-motion and partly on environmental cues \cite{Moser2008}.   Locally within the MEC, grid cells share the same orientation and periodicity, but vary randomly in phase\cite{Hafting2005}. The scale of grid fields varies systematically along the dorso-ventral axis of the MEC (Fig.~1A)\cite{Hafting2005,Stensola2012}.

How does the  grid system represent spatial location and what function does  the striking triangular lattice organization and systematic variation in grid scale serve?    Here, we begin by assuming that grid cell scales are organized into discrete modules \cite{Stensola2012}, and propose that the grid system follows a principle of economy by minimizing the number of neurons required to achieve a given spatial resolution. Our hypothesis, together with general assumptions about tuning curve shape and decoding mechanism, predicts a geometric progression of grid scales. %the grid system implements a two-dimensional analog of a base-b number system. 
The theory further determines the mean ratio between scales, explains the triangular lattice structure of grid cell firing maps, and makes several additional predictions that can be subjected to direct experimental test.     For example, the theory predicts that the ratio of adjacent grid scales will be modestly variable within and between animals  with a  mean in the range $1.4 - 1.7$ depending on the assumed decoding mechanism used by the brain.  This prediction is quantitatively supported by recent experiments \cite{Barry2007,Stensola2012}. In a simple decoding scheme, the scale ratio in an $n$-dimensional environment is predicted to be close to $\sqrt[n]{e}$. We also estimate the total number of scales providing the spatial resolution necessary to support navigation over typical behavioral distances, and show that it compares favorably with estimates from recent experimental measurements \cite{Stensola2012}.
%
%We will find that that the optimal ratio of grid scales is close to $\sqrt[n]{e}$ in $n$-dimensions, for a wide variety of tuning curve shapes and decoding strategies. We will see that the minimum is shallow, so that modest variations around optimality increase neuron number by only a small amount.  This implies that the ratios of adjacent grid scales will be somewhat variable around a fixed mean within and between animals.    The predictions of different specific models agree with each other, and with experimental data, within this variability. 

%Figure1
\begin{figure}
\centering
\includegraphics[keepaspectratio,width=0.9\linewidth]{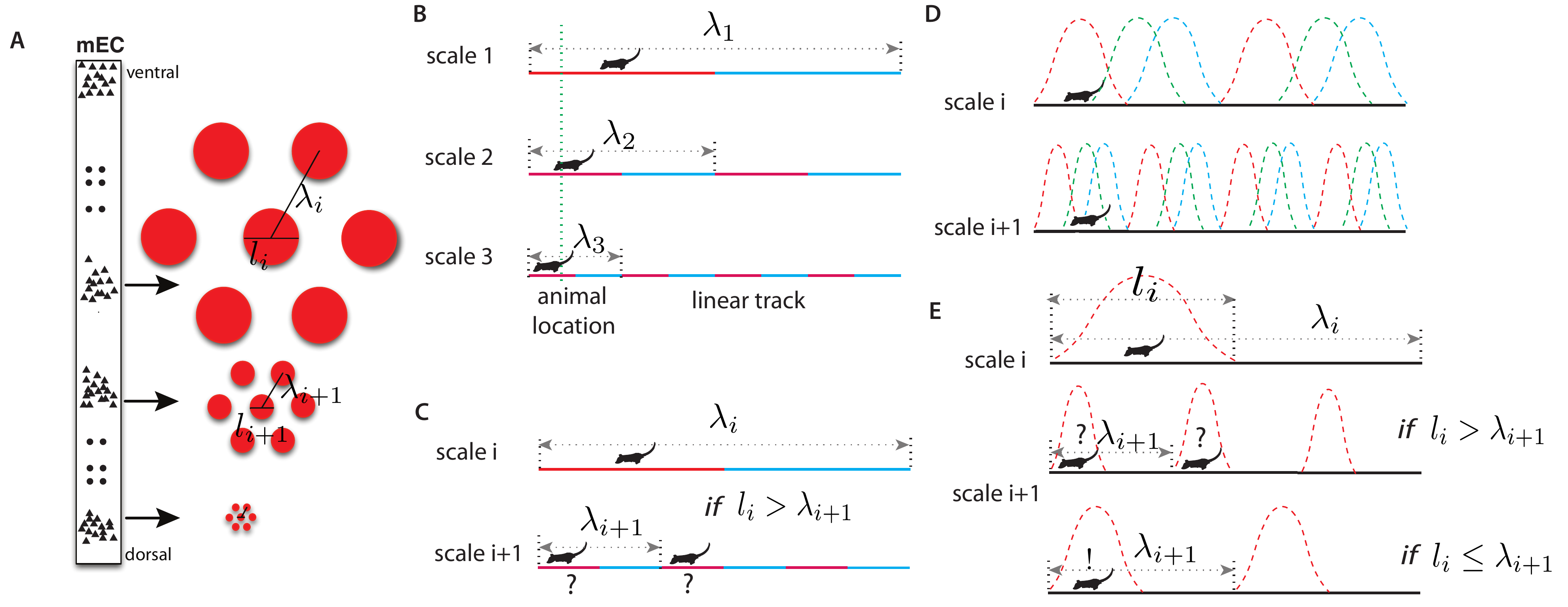}

\caption{Representing place in the grid system.    
{\bf (A)} Grid cells (small triangles) in the medial entorhinal cortex (MEC) respond when the animal is in a triangular lattice of physical locations (red circles; sometimes also called a ``hexagonal lattice'')  \cite{Fyhn2004,Hafting2005}.     The scale of periodicity (the ``grid scale'', $\lambda_i$) and the size of the regions evoking a response (the ``grid field width'', $l_i$) vary systematically along the dorso-ventral axis of the MEC \cite{Hafting2005}.    
{\bf (B)} A simplified binary grid scheme for encoding location along a linear track.   At each scale ($\lambda_i$) there are two grid cells (red vs. blue firing fields).  The periodicity and grid field widths are halved at each successive scale.  
{\bf (C)} Decoding is ambiguous if the grid field width at scale $i$ exceeds the grid periodicity at scale $i+1$.  E.g., if the grid fields marked in red respond at scales $i$ and $i+1$, the animal might be in either of the two marked locations. 
{\bf (D)} We extend the binary code of panel \emph{B} to the more realistic case of populations of noisy neurons with overlapping tuning curves. 
{\bf (E)} The relationship between grid periodicity, $\lambda_i$, and grid field width, $l_{i}$. In the winner-take-all case, decoded position will be ambiguous unless $l_i \le \lambda_{i+1}$, analogously to the situation depicted in panel \emph{C}.}    
\label{fig:grid}
\end{figure}

\section*{Results}
\begin{figure}
\centering		
\includegraphics[keepaspectratio,width=0.8\linewidth]{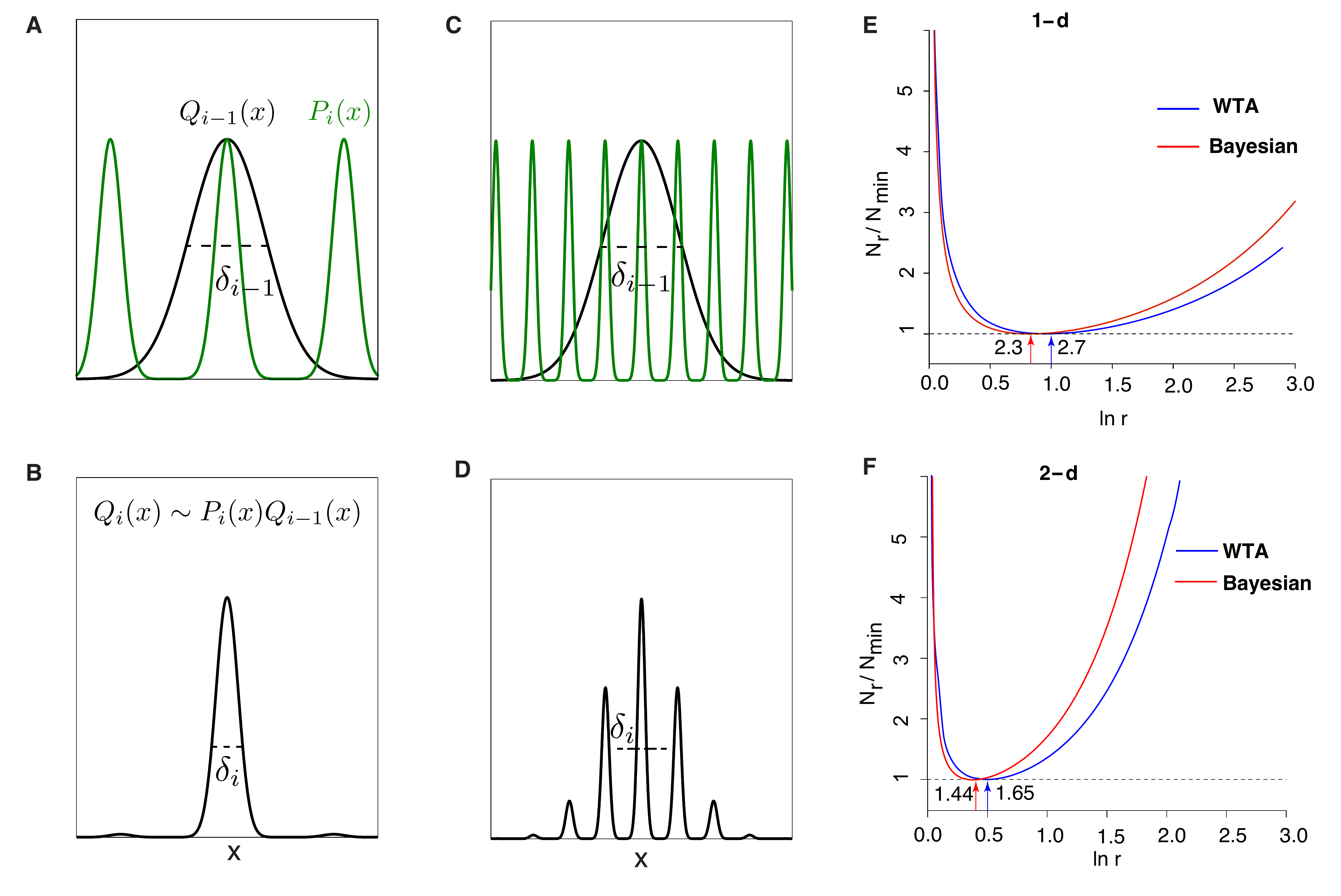}
\hspace{0.03\linewidth}
\caption{
{\bf (A-D)} Trade-off between precision and ambiguity in the Bayesian decoder. {\bf(A)} Information about position given the responses of all grid cells at scales smaller than module $i$ is summarized by the posterior $Q_{i-1}(x)$ (black curve), and the uncertainty in position is given by the standard deviation $\delta_{i-1}$. Grid cells in module $i$ contribute the periodic posterior $P_i(x)$ (green curve). 
{\bf(B)} The updated posterior combining module $i$ with all larger-scale modules is given by the product $Q_i(x) \sim P_i(x)Q_{i-1}(x)$, and has the reduced uncertainty $\delta_i$.  
{\bf(C)} Precision is improved by increasing the scale factor, thereby narrowing the peaks of $P_i(x)$. However, the periodicity shrinks as well, increasing ambiguity. 
{\bf(D)} Posterior $Q_i(x)$ given by combining the modules shown in \emph{C}. Ambiguity from the secondary peaks leads to an overall uncertainty $\delta_i$ larger than in \emph{B}, despite the improved precision from the narrower central peak. There is thus an optimal scale factor somewhere between that in \emph{A}, \emph{B} and in \emph{C}, \emph{D}.
{\bf (E)} The optimal ratio $r$ between adjacent scales in a hierarchical grid system in one dimension for a simple winner-take-all decoding model (blue curve, WTA) and a Bayesian decoder (red curve). Here $N_r $ is the number of neurons required to represent space with resolution $R$ given a scaling ratio $r$, and $N_{\rm min}$ is the number of neurons required at the optimum.  In both decoding models, the ratio $N_r / N_{\rm min}$ is independent of resolution, $R$. For the winner-take-all model, $N_r \propto r / \ln r$, as derived in the main text, and the curve for the Bayesian model is derived numerically as described in Supplemental Sec.~5. The winner-take-all model predicts that the minimum number of neurons is achieved for $r = e \approx 2.7$, while the Bayesian decoder predicts $r \approx 2.3$.   The minima of the two curves lie within each others' shallow basins. 
{\bf (F)} Same as \emph{E}, but in two dimensions with a triangular grid. The winner-take-all curve in this case is $N_r \propto r^2 / \ln (r^2)$ (see main text), and the minima occur at $r = \sqrt{e} \approx 1.65$ for winner-take-all and $r \approx 1.44$ for the Bayesian case.  The shallowness of the basins around these minima predicts that some variability of adjacent scale ratios is tolerable, both within and between animals. }
\label{fig:scaling}
\end{figure}

\subsection*{General grid coding in one dimension}
Consider a one dimensional grid system that develops when an animal runs on a linear track.  Suppose that grid fields develop at a discrete set of periodicities $\lambda_1 > \lambda_2 > \cdots > \lambda_m$ (Fig.~1A).  We will refer to the population of grid cells sharing each periodicity $\lambda_i$ as one module. It will prove convenient to define ``scale factors''  $r_i \equiv \frac{\lambda_i}{\lambda_{i+1}}$. Here  $\lambda_1$ could be the length of the entire track and we do not assume any further relation between the $\lambda_i$, such as a common scale ratio (i.e., in general $r_1 \ne r_2 \ne \cdots \ne r_{m-1}$).  Now let the widths of grid fields in each module be denoted $l_1, l_2, \cdots l_m$.     Within any module, grid cells have a variety of spatial phases so that at least one cell may respond at any physical location (Fig.~1D). To give uniform coverage of space, the  number of grid cells $n_i$ at scale $i$ should be proportional to $\lambda_i / l_i$  -- thus we write  $ n_i = d \lambda_i/l_i $ in terms of a ``coverage factor'' $d$ that represents the number of grid fields overlapping each point in space.  We  assume that $d$ is the same at each scale.   In terms of these parameters, the total number of grid cells is $N = \sum_{i=1}^m n_i = \sum_{i=1}^{m} d \frac{\lambda_i}{l_i}$.

Grid cells with smaller scales provide more local spatial information than those with larger scales, owing to their smaller $l_i$. However, this increased resolution comes at a cost: the smaller periodicity $\lambda_i$ of these cells leads to increased ambiguity (Fig.~1C,E, Fig.~2A-D).  In this paper, we study coding schemes in which information from grid cells with larger scales is used to resolve this ambiguity in the smaller scales, while the smaller scales provide improved local resolution (Fig. ~1E). In such a system, resolution may thus be improved by increasing the total number of modules $m$. Alternatively, the field widths $l_i$ may be made smaller relative to the periodicities $\lambda_i$; however, this necessitates using more neurons at each scale in order to maintain the same coverage $d$. Improving resolution by either mode therefore requires additional neurons. An efficient grid system will minimize the number of grid cells providing a fixed resolution $R$; we shall demonstrate how the parameters of the grid system, $r_i$, $l_i / \lambda_i$, and $m$, should be chosen to achieve this optimal coding. We will characterize efficient grid systems in the context of two decoding methods at extremes of complexity.

We first consider a decoder which consider the animal as localized within the grid field of the most responsive cell in each module \cite{WinnerTakeAll1,WinnerTakeAll2}.  Such a  ``winner-take-all'' scheme is at one extreme of decoding complexity and could be easily implemented by neural circuits.     Any decoder will have to threshold grid cell responses at the background noise level, so that the firing fields are effectively compact (Fig.~1D). Grid cell recordings suggest that the firing fields are, indeed, compact \cite{Hafting2005}.   The uncertainty in the animal's location at grid scale $i$ is given by the grid field width  $l_i$.   The smallest scale that can be resolved in this way is $l_m$, we therefore define the resolution of the grid system as the ratio of the largest to the smallest scale, $R_1 = \lambda_1/l_m$.   In terms of  scale factors  $r_i \equiv \frac{\lambda_i}{\lambda_{i+1}}$, we can write the resolution as  $R_1  = \prod_{i=1}^{m} r_i$, where we also defined $r_m \equiv \frac{\lambda_m}{l_m}$.  Unambiguous decoding requires that  $l_i \le \lambda_{i+1}$ (Fig.~1C,E), or, equivalently, $\frac{\lambda_i}{l_i} \ge r_i$.   To minimize $N = d \sum_i \lambda_i / l_i$, all the $\frac{\lambda_i}{l_i}$  should be as small as possible; so this fixes $\frac{\lambda_i}{l_i} = r_i$. 
 Thus we are reduced to minimizing the sum $N = d \sum_{i=1}^{m} r_i$  over the parameters $r_i$, while fixing the product $R = \prod_i r_i$.   Because this problem is symmetric under permutation of the indices $i$, the optimal $r_i$ turn out to all be equal, allowing us to set $r_i = r$ (Supplementary Material). Our optimization principle thus predicts a common scale ratio, giving a geometric progression of grid periodicities. The constraint on resolution then gives $m = \log_r R$, so that we seek to minimize $N(r) = d \, r \log_r R$ with respect to $r$: the solution is $r = e$ (Fig.~2E; details in Supplementary Material).   Therefore, for each scale $i$, $\lambda_{i} = e \, \lambda_{i+1}$ and $\lambda_{i} = e \, l_i$. Here we treated $N$ and $m$ as continuous variables -- treating them as integers throughout leads to the same result through a more involved argument (Supplementary Material).       The coverage factor $d$ and the resolution $R$ do not appear in the optimal ratio of scales.

The brain might implement the simple decoding scheme above via a winner-take-all mechanism \cite{WinnerTakeAll1,WinnerTakeAll2,WinnerTakeAll3}.    But the brain is also capable of implementing far more complex decoders.  Hence, we also consider a Bayesian decoding scheme that optimally combines information from all grid modules.    In such a setting, an ideal decoder should construct the posterior probability distribution of the animal's location given the noisy responses of all grid cells.  The population response at each scale $i$ will give rise to a posterior over location $P(x | i)$, which will have the same periodicity $\lambda_i$ as the individual grid cells' firing rates (Fig.~2A). The posterior given all $m$ scales, $Q_m(x)$, will be given by the product $Q_m(x) = {\cal N} \, \Pi_{i=1}^m P(x|i)$, assuming independent response noise across scales (Fig.~2B). Here ${\cal N}$ is a normalization factor. The animal's overall uncertainty about its position will then be related to the standard deviation $\delta_m$ of $Q_m(x)$, we therefore quantify resolution as $R = \lambda_1 / \delta_m$. $\delta_m$, and therefore $R$, will be a function of all the grid parameters (Supplementary Material). In this framework, ambiguity from too-small periodicity $\lambda_i$ decreases resolution, as does imprecision from too-large field width $l_i$. We thus need not impose an a priori constraint on the minimum value of $\lambda_i$, as we did in the winner-take-all case: minimizing neuron number while fixing resolution automatically resolves the tradeoff between precision and ambiguity (Fig.~2A-D). To calculate the resolution explicitly, we note that when the coverage factor $d$ is very large, the distributions $P(x|i)$ will be well-approximated by periodic arrays of Gaussians (even though individual tuning curves need not be Gaussian). We can then minimize the neuron number, fixing resolution, to obtain the optimal scale factor $r \approx 2.3$: slightly smaller than, but close to the winner-take-all value, $e$ (Fig.~2E; details in Supplementary Material). As before, the optimal scale factors are all equal so we again predict a geometric progression of scales. 

%Consider grid tuning curves that have non-zero (but possibly very small) firing rates throughout space. Provided the coverage factor $d$ is large enough, the posterior distribution of position conditioned on the responses of grid cells at scale $i$, $P(x|i)$, will be well-approximated by a periodic array of Gaussians.  Note that we are \emph{not} assuming that the tuning curves themselves are Gaussian: only that firing rate is everywhere non-zero and that coverage is dense. (If the grid fields were compact, as assumed previously, $P(x\,|\,i)$ could instead be approximated by truncated Gaussians). The standard deviation of each Gaussian ($\sigma_i$) will be a  function of the coverage factor $d$ and the widths of individual grid fields $l_i$.    The posterior distribution given the responses of {\it all} $m$ scales will be a product of the $P(x|i)$: $Q_m(x) = {\cal N} \, \Pi_{i=1}^m P(x|i)$, where ${\cal N}$ is a normalization factor.  $Q_m(x)$ will also be well described by a Gaussian.     We may then quantify the  resolution of the system as the ratio of  the largest scale (e.g., the track length) to the standard deviation of $Q_m$.    The optimal scaling ratio $r = \lambda_i/\lambda_{i+1}$, derived by minimizing the number of cells needed to achieve the desired resolution, is $r = 2.3$, slightly smaller than, but close to, $e$ (Fig.~2B; details in Supplementary Material). 

It is apparent from Fig.~2E that the minima for both the Bayesian decoder and the winner-take-all decoder are  shallow, so that the scaling ratio $r$ may lie anywhere within a basin around the optimum at the cost of a small number of additional neurons. Even though our two decoding strategies lie at extremes of complexity (one relying just on the most active cell at each scale and another optimally pooling information in the grid population) their respective ``optimal intervals'' substantially overlap. 
%For the simple winner-take-all model, a scale factor lying in the interval $(2.05, 3.84)$ would increase the required number of neurons by no more than $5\%$ of the minimum. For the Bayesian decoding model, the corresponding interval is $(1.78, 3.11)$, which has substantial overlap with the interval predicted by the winner-take-all model. 
That these two very different models make overlapping predictions suggests that our theory is robust to variations in the detailed shape of grid cells' grid fields and the precise decoding model used to read their responses. Moreover, such considerations also suggest that these coding schemes have the capacity to tolerate developmental noise: different animals could develop grid systems with slightly different scaling ratios, without suffering a large loss in efficiency.

%Figure3
\begin{figure}
\centering
\includegraphics[keepaspectratio,width=0.8\linewidth]{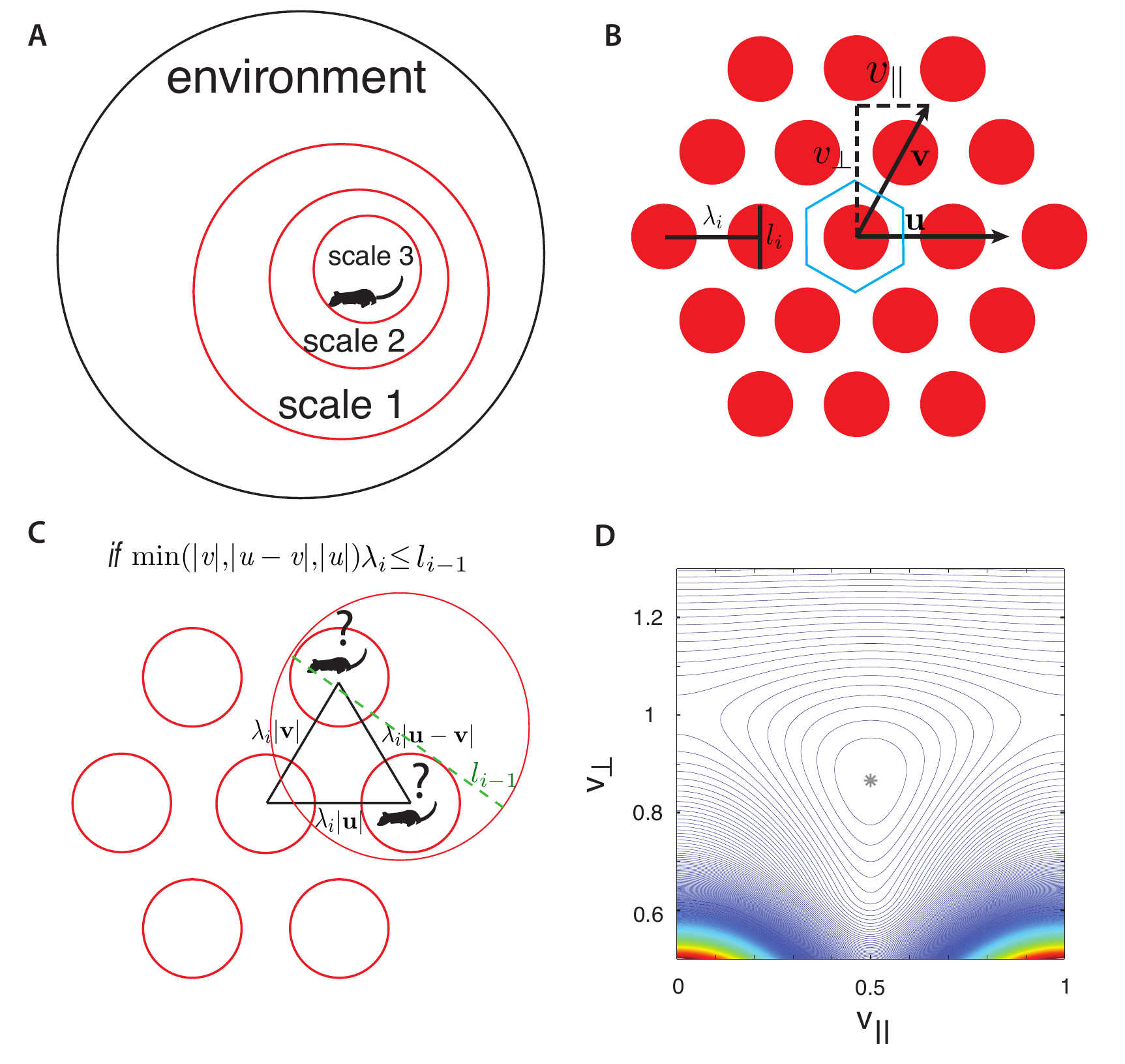}
\caption{ 
{\bf (A)} Two dimensional analog of a grid scheme with circular firing fields.  
{\bf (B)} A general two-dimensional lattice may be parameterized by two vectors $\mathbf{u}$ and $\mathbf{v}$ and a periodicity parameter $\lambda_i$. We take $\mathbf{u}$ to be a unit vector, so that the spacing between peaks along the $\mathbf{u}$ direction is $\lambda_i$, and denote the two components of $\mathbf{v}$ by $v_\parallel$, $v_\perp$.   The blue-bordered region is a fundamental domain of the lattice, the largest spatial region that may be unambiguously represented. 
{\bf (C)}  The two dimensional analog of the ambiguity in Fig.~1C, E for the winner-take-all decoder.  If the grid fields in scale $i$ are too close to each other relative to the size of the grid field of scale $i-1$ (i.e. $l_{i-1}$), the animal might be in one of several locations.
{\bf (D)} Contour plot of normalized neuron number $N / N_{\rm min}$ in the Bayesian decoder, as a function of the grid geometry parameters $v_\perp, v_\parallel$ after minimizing over the scale factors for fixed resolution $R$. As in Fig.~2E,F, the normalized neuron number is independent of $R$. The spacing between contours is 0.01, and the asterisk labels the minimum at $v_\parallel = 1/2, v_\perp = \sqrt{3}/2$; this corresponds to the triangular lattice. }
\label{fig:q}
\end{figure}

\subsection*{General grid coding in two dimensions}
How do these results extend to two dimensions?       Let $\lambda_i$ be the distance between neighboring peaks of grid fields of width $l_i$ (Fig.~1A).  Assume in addition that a given cell responds on a lattice whose vertices are located at the points $\lambda_i (n\mathbf{u} + m\mathbf{v})$, where $n, m$ are integers and $\mathbf{u}, \mathbf{v}$ are linearly independent vectors generating the lattice (Fig.~3B). We may take $\mathbf{u}$ to have unit length ($|\mathbf{u}| = 1$) without loss of generality, however $|\mathbf{v}| \ne 1$ in general. It will prove convenient to denote the components of $\mathbf{v}$ parallel and perpendicular to $\mathbf{u}$ by $v_\parallel$ and $v_\perp$, respectively (Fig.~3B). The two numbers $v_\parallel, v_\perp$ quantify the geometry of the grid and are additional parameters that we may optimize over: this is a primary difference from the one-dimensional case. We will assume that $v_\parallel$ and $v_\perp$ are independent of scale; this still allows for relative rotation between grids at different scales.
%that lies at the vertices of a regular tiling of the plane.   There are exactly three possible choices of such grids  (Fig.~3).    A

At each scale, grid cells have different phases so that at least one cell responds at each physical location.   The minimal number of phases required to cover space is computed by dividing the area of the unit cell of the grid ($\lambda_i^2 ||\mathbf{u} \times \mathbf{v}|| = \lambda_i^2 |v_\perp|$) by the area of the grid field.  As in the one-dimensional case, we define a coverage factor $d$ as the number of neurons covering each point in space, giving for the total number of neurons $N = d |v_\perp| \sum_i (\lambda_i / l_i)^2$. 
%This gives $g \, (\lambda_i/l_i)^2$ where $g$ is a geometrical factor that depends on the particular grid.   We will compute $g$ explicitly below, but leave it abstract for now.    The number of neurons at each scale is $n_i =  d \, g \, (\lambda_i / l_i)^2$, including, as in the one-dimensional case, a  ``coverage factor'' $d$ parametrizing any excess beyond the minimum required to cover space.   The total number of neurons is $N =  d \, g \sum_i (\lambda_i / l_i)^2$.   

As before,  consider a simple model where grid fields lie completely within compact regions and assume a decoder which selects the most activated cell  \cite{WinnerTakeAll1,WinnerTakeAll2,WinnerTakeAll3}.  In such a model, each scale $i$ serves to localize the animal within a circle of diameter $l_i$.   The spatial resolution  is summarized by the square of the ratio of the largest scale $\lambda_1$ to the smallest scale $l_m$: $R_2 = (\lambda_1 / l_m)^2$.  In terms of the scale factors $\tilde{r}_i = \lambda_i / \lambda_{i+1}$ we write $R_2  = \prod_{i=1}^m \tilde{r}_i^2$, where we also define $\tilde{r}_m = \lambda_m / l_m$. To decode the position of an animal unambiguously, each cell at scale $i$ should have at most one grid field within a region of diameter $l_{i-1}$.  Since the nearest firing fields lie at a distance $\lambda_{i}$ along the three grid axes $\mathbf{u}, \mathbf{v}$, and $\mathbf{u}-\mathbf{v}$, we require  $\min(|\mathbf{v}|, |\mathbf{u}-\mathbf{v}|, 1) \cdot \lambda_{i} \ge l_{i-1}$ in order to avoid ambiguity (Fig.~3C).
%    This requires that $l_i \leq \lambda_{i+1}$ or, equivalently, $\lambda_i/l_i \geq \tilde{r}_i$  
To minimize $N$ we must make $\lambda_{i-1} / l_{i-1} = \tilde{r}_{i-1} \lambda_{i} / l_{i-1}$ as small as possible, so that $\lambda_{i} = l_{i-1}$, which is only possible if $|\mathbf{v}| \ge 1, |\mathbf{u}-\mathbf{v}| \ge 1$. We then have $N = d |v_\perp| \sum_i \tilde{r}_i^2$.   We now seek parameters $v_\parallel, v_\perp, \tilde{r}_i$ that minimize $N$ while fixing the resolution $R_2$. Since $R_2$ does not depend on the geometric parameters $v_\parallel, v_\perp$, we may determine these parameters by simply minimizing $N$, which is equivalent to minimizing $|v_\perp|$ subject to the constraints $|\mathbf{v}| \ge 1, |\mathbf{u}-\mathbf{v}| \ge 1$. This optimization picks out the triangular lattice with $v_\perp = \sqrt{3}/2, v_\parallel = 1/2$. Note that this formulation is mathematically analogous to the optimal sphere-packing problem, for which the solution in two dimensions is also the triangular lattice \cite{Thue}. As for the scale factors $\tilde{r}_i$, the optimization problem is mathematically the same as in one dimension if we formally set $r_i \equiv \tilde{r}_i^2$.  This gives the optimal ratio $\tilde{r}_i^2 = e$ for all $i$ (Fig.~2F).   We conclude that in two dimensions, the optimal ratio of neighboring grid periodicities is $\sqrt{e} \approx 1.65$ for the simple winner-take-all decoding model, and the optimal lattice is triangular. 

  The Bayesian decoding model  can also be extended to two dimensions with the posterior distributions $P(x|i)$ becoming sums of Gaussians with peaks on the two-dimensional lattice.  In analogy with the one-dimensional case, we then derive a formula for the resolution $R_2 = \lambda_1 / \delta_m$ in terms of the standard deviation $\delta_m$ of the posterior given all scales. $\delta_m$ may be explicitly calculated as a function of the scale factors $\tilde{r}_i$ and the geometric factors $v_\parallel, v_\perp$, and the minimization of neuron number may then be carried out numerically (Supplementary Material). In this approach the optimal scale factor turns out to be $\tilde{r}_i \approx 1.4$ (Fig.~2F), and the optimal lattice is again triangular (Fig.~3D). 

Once again, the optimal scale factors in both decoding approaches lie within overlapping shallow basins, indicating that our proposal is robust to variations in grid field shape and to the precise decoding algorithm (Fig.~2F).  In two dimensions, the required neuron number will be no more than $5\%$ of the minimum if the scale factor is within $(1.43, 1.96)$ for the winner-take-all model and $(1.28, 1.66)$ for the Bayesian model.  These ``optimal intervals"  are narrower than in the one-dimensional case, and have substantial overlap.

The fact that both of our decoding models predicted the triangular lattice as optimal is a consequence of the fact that they share a very general symmetry. The resolution formula in both problems is invariant under a common rotation and a common rescaling of all firing rate maps. The neuron number shares this symmetry, as well. The rotation invariance  implies that the resolution only depends on grid geometry through $v_\perp, v_\parallel$, and the rescaling invariance implies that it only depends on $\lambda_i, l_i$ through the dimensionless ratios $r_i, \lambda_i / l_i$. However, even after restricting the parameters in this way, the rotation- and rescaling-invariance has a nontrivial consequence. The transformation $v_\perp \to -v_\perp / |\mathbf{v}|^2, v_\parallel \to v_\parallel / |\mathbf{v}|^2, l_i \to l_i / |\mathbf{v}|$ can be seen to be equivalent to a rotation of the grid combined with a scaling by $|\mathbf{v}|$ (Supplementary Material), and therefore must leave the resolution and neuron number invariant. If there is a unique optimal grid, it must then also be invariant under this transformation: this constraint is only satisfied by the square grid ($v_\perp = 1, v_\parallel = 0$) and the triangular grid ($v_\perp = \sqrt{3}/2, v_\parallel = 1/2$). Between these two, the triangular grid has the smaller $v_\perp$ and so will minimize neuron number (see Supplementary Material for a more rigorous discussion). We therefore see that the optimality of the triangular lattice is a very general consequence of minimizing neuron number for fixed resolution, and expect the result to hold for a wide range of decoders.  

%Figure4
\begin{figure}
\centering
\includegraphics[keepaspectratio,width=0.7\linewidth]{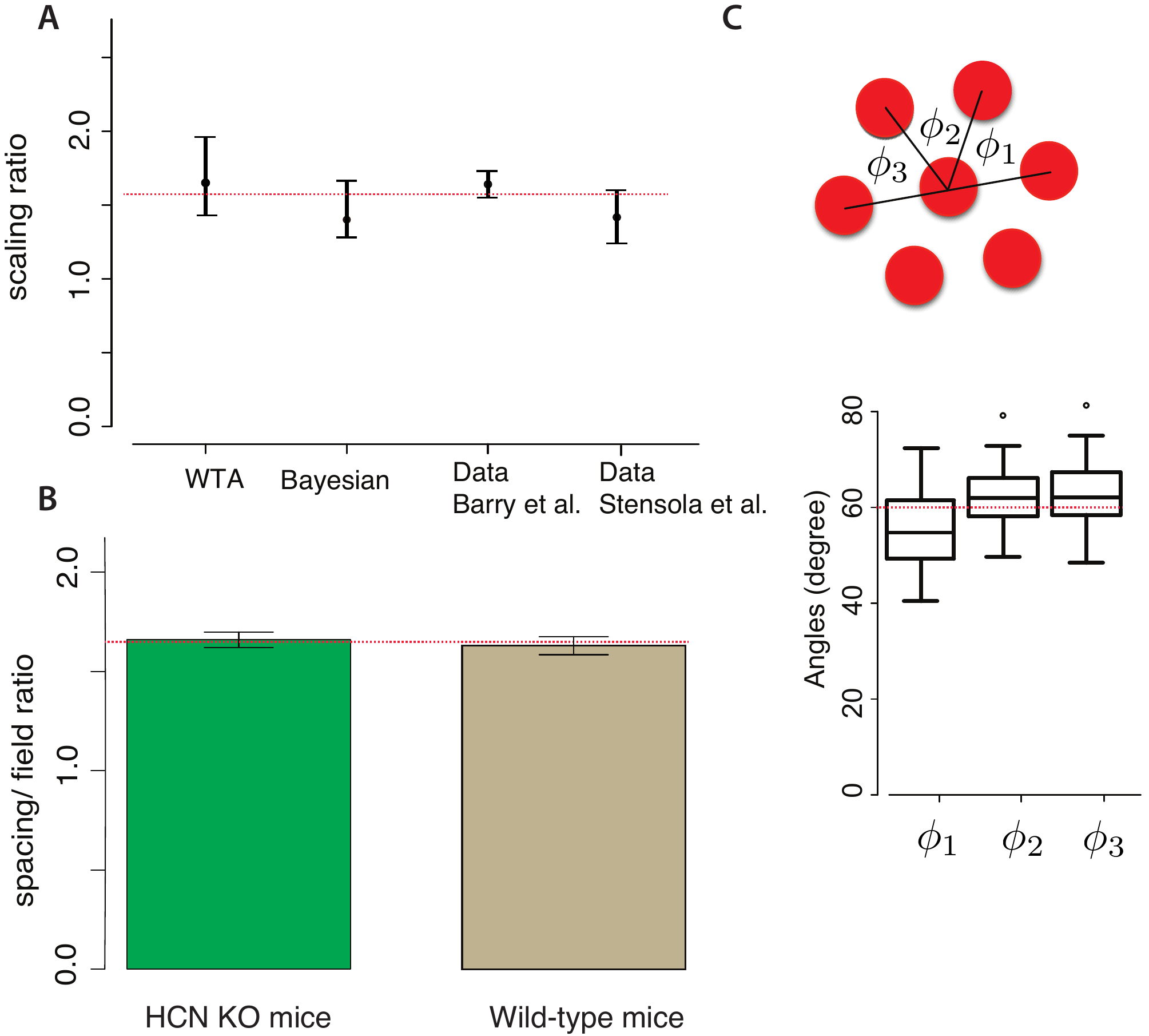}

\caption{
{\bf (A)} Our models predict grid scaling ratios that are consistent with experiment.  `WTA' (Winner-Take-All) and `Bayesian' represent  predictions from two decoding models; the dot is the scaling ratio minimizing neuron number and the error bars represent the interval within which the neuron number will be no more than $5 \%$ higher than the minimum. For the experimental data, the dot represents the mean measured scale ratio and the error bars represent $\pm$ one standard deviation.  Data were replotted from \cite{Barry2007,Stensola2012}.  The dashed red line shows a consensus value running through the two theoretical predictions and  the two experimental  datasets. 
{\bf (B)} The mean ratio between grid periodicity ($\lambda_i$)  and the diameter of grid fields ($l_i$) in mice (replotted from  \cite{Giocomo2011}). Error bars indicate $\pm$ one S.E.M. For both wild type mice and HCN knockouts (which have larger grid periodicities) the ratio is consistent with $\sqrt{e}$ (dashed red line).  
{\bf (C)}  The response lattice of grid cells in rats forms an equilateral triangular lattice with $60^\circ$ angles between adjacent lattice edges  (replotted from \cite{Hafting2005}, $n=45$ neurons from 6 rats).  Dots represent the outliers. 
 }
\label{fig:nn}
\end{figure}

\subsection*{Comparison to experiment}
Our predictions agree with experiment \cite{Barry2007,Giocomo2011,Stensola2012} (see Supplementary Material for details of the data re-analysis).  Specifically, Barry et al., 2007 (Fig.~4A) reported the grid periodicities measured at three locations along the dorso-ventral axis of of the MEC in rats and found ratios of $\sim 1$, $\sim 1.7$ and $\sim 2.5 \approx 1.6 \times 1.6$ relative to the smallest period\cite{Barry2007} .   The ratios of adjacent scales reported in \cite{Barry2007} had a mean of $1.64 \pm 0.09$ (mean $\pm$ std. dev., $n=6$), which almost precisely matches the mean scale factor of $\sqrt{e}$ predicted from the winner-take-all decoding model, and is also consistent with the Bayesian decoding model.   Recent analysis based on larger data set \cite{Stensola2012} confirms the geometric progression of the grid scales. The mean adjacent scale ratio is $1.42 \pm 0.17$ (mean $\pm$ std. dev., $n=24$) in that data set, accompanied by modest variability of the scaling factors both within and between animals.  These measurements again match  both our models (Fig.~4A). The optimal grid was triangular in both of our models, this again matches measurements (Fig.~4C) \cite{Hafting2005,Moser2008,Stensola2012}. 

The winner-take-all model  also predicts the ratio between grid period and grid field width:  $\lambda_i / l_ i = \lambda_i/ \lambda_{i+1} =  \sqrt{e} \approx 1.65$.    A recent study measured the ratio between grid periodicity and grid field size to be $1.63 \pm 0.035$ (mean $\pm$ S.E.M., $n=48$) in wild type mice\cite{Giocomo2011}, consistent with our predictions (Fig.~4B).  This ratio was unchanged, $1.66 \pm 0.03$ (mean $\pm$ S.E.M., $n=86$), in HCN1 knockout strains whose absolute grid periodicities increased relative to the wild type\cite{Giocomo2011}.  The Bayesian model does not make a direct prediction about grid field width; it instead works with the standard deviation of the posterior $P(x\,|\,i)$, $\sigma_i$ (Supplementary Material). This parameter is predicted to be $\sigma_i = 0.19\lambda_i$ in two dimensions, but cannot be directly measured from data. It is related to the field width $l_i$ by a proportionality factor whose value depends on detailed tuning curve shape, noise properties, firing rate, and firing field density (Supplementary Material). 

We can estimate the total number of modules, $m$, by estimating the requisite resolution $R_2$ and using the relationship $m = \log{R_2} / \log{\tilde{r}^2}$. Assuming that the animal must be able to navigate an environment of area $\sim (10\, \mathrm{m})^2$, with a positional accuracy on the scale of the rat's body size, $\sim(10\, \mathrm{cm})^2$, we get a resolution of $R_2 \sim 10^4$. Together with the predicted two-dimensional scale factor $\tilde{r}$, this gives $m \approx 10$ as an order-of-magnitude estimate. Indeed, in \cite{Stensola2012}, 4-5 modules were discovered in recordings spanning up to 50\% of the dorsoventral extent of MEC; extrapolation gives a total module number consistent with our estimate.

%
%
%The remaining variable in our analysis is the grid geometry (Fig.~3C-E).  The number of neurons associated to a these grids $N = g \, d \sum_i \tilde{r}_i^2$, where the geometrical factor $g$ depends on the grid type.  This factor (times $\tilde{r}_i^2$) essentially measures the minimal number of phases (i.e. different neurons) required to cover all the space within a unit cell of the grid.   It is computed as the ratio of the unit cell area to the individual grid field area (see Supplementary Material), giving  $g_{{\rm triang.}}= {2 \sqrt{3} \over \pi} \ <  \ g_{{\rm square}} = {4 \over \pi} \ <  \ g_{{\rm hex.}} = {3 \sqrt{3} \over \pi}$.  Thus, minimizing the number of neurons selects the triangular grid.   This again matches measurements (Fig.~4C) \cite{Hafting2005,Moser2008}. It is worth noting that a more general argument for the optimality of the triangular lattice can be found in the study of circular packings \cite{Thue}. 
%

\section*{Discussion}
We have shown that a grid system with a discrete set of periodicities, as found in the entorhinal cortex, should use a common scale factor $r$ between modules to represent spatial location with the fewest neurons. In one dimension, this organization may be thought of intuitively as implementing a neural analog of a base-$b$ number system. Each scale localizes the animal to some coarse region of the environment, and the next scale subdivides that region into $b = r$ ``bins'' (Fig.~1C). Our problem of minimizing neuron number while fixing resolution is analogous to minimizing the number of symbols needed to represent a given range $R$ of numbers in a base-$b$ number system. Specifically, $b$ symbols are required at each of $\log_b R$ positions, and minimizing the total, $b \log_b R$, with respect to $b$ gives an optimal base $b = e$.  Our full theory can thus be seen as a generalization of this simple fixed-base representational scheme to noisy neurons encoding two-dimensional location.

The existing data agree with our predictions for the ratios of adjacent scales  within the variability tolerated by our models (Fig.~4).   Further tests of our theory are possible.    For example, a direct generalization of our reasoning says that in n-dimensions the optimal ratio between grid scales will be near $\sqrt[n]{e}$, with $n = 3$ having possible relevance to the grid system \cite{Hayman2011} in, e.g., bats\cite{Yartsev2011}.    In general, the theory can be tested by comprehensive population recordings of grid cells along the dorso-ventral axis for animals moving in one, two and three dimensional environments.     There is some evidence that humans also have a grid system \cite{Doeller 2010}, in which case our theory may have relevance to the human sense of place.

We assumed that the grid system should minimize the number of neurons required to achieve a given spatial resolution.   
 In fact, any cost which increases monotonically with the number of neurons would lead to the same optimum.   
Of course,  completely different proposals for the functional architecture of the grid system \cite{Fiete2008,Fiete2011,Chuck}and associated cost functions will lead to different predictions.    For example, \cite{Fiete2008,Fiete2011} showed that a grid  implementing a ``residue number system'' (in which adjacent grid scales should be relatively prime) will maximize the range of positions that can be encoded.  This theory makes distinct predictions for the ratios of adjacent scales (the different periods are relatively prime) and, in its original form,  predicts neither  the ratio of grid field width to periodicity nor the organization in higher dimensions, except perhaps by interpreting higher dimensional grid fields as a product of one-dimensional fields.   The essential difference between these two theories lies in the fundamental assumptions:   we minimize the number of neurons needed to represent space with a given resolution and range, as opposed to maximizing the range of locations that may be uniquely encoded.     

Grid coding schemes represent position more accurately than place cell codes given a fixed number of neurons \cite{mathis2012,mathisPRL}.    Furthermore,  in one dimension a geometric progression of grids that are self-similar at each scale minimizes the asymptotic error in recovering an animal's location given a fixed number of neurons \cite{mathis2012}.     The two dimensional grid schemes discussed in this paper will share the same virtue.

 The scheme that we propose may also be more developmentally plausible, as each scale is determined by applying a fixed rule (rescaling by $r$) to the anatomically adjacent scale. This could be encoded, for example, by a morphogen with an exponentially decaying concentration gradient along the dorsoventral axis, something readily attainable in standard models of development. This differs from the global constraint that all scales be relatively prime, for which the existence of a local developmental rule is less plausible.   As we showed, the location coding scheme that we have described is also robust to variations in the precise value of the scale ratio $r$, and so would tolerate variability within and between animals. 
  
%  
%In \cite{Stensola2012}, an average grid scale ratio of $1.42$ (see Fig.~4) was measured over about $5$  scales  in a recording environment with a 2m diameter and relatively little fine-grained structure.     Thus the smallest scale represented was about 30 cm.     In our formalism, the relationship between resolution, number of scales, and scale factor is given by $R = r^m$.  An interesting question for the future is whether the number  of grid scales, and their structure, varies when the recording environment is larger (e.g. a basketball court), or has ``local texture'' below the $\sim 30$~cm scale.       Large-scale population recordings along the dorso-ventral axis of entorhinal cortex when rodents navigate large, more finely structured environments will definitively decide between our theory and the alternatives  \cite{Fiete2008,Fiete2011,Chuck}.

\small{\bibliography{scibib}

\bibliographystyle{Science}

\begin{scilastnote}
\item NSF grants PHY-1058202, EF-0928048  and PHY-1066293 supported this work, which was completed at the Aspen Center for Physics.   VB was also supported by the Fondation Pierre Gilles de Gennes.  JP was supported by the C.V. Starr Foundation. While this paper was being written we became aware of related work in progress by Charles Stevens and Trygve Solstad (personal communication).
\end{scilastnote}

\clearpage

\begin{center}
\Large{ {\bf Supplementary~materials}}
\end{center}

\section{Optimizing a ``base-b'' representation of one-dimensional space}
\label{s:baseb}
Suppose that we want to resolve location with a precision $l$ in a track of length $L$.   In terms of the resolution $R = L/l$, we have argued in the discussion of the main text that a ``base-b'' hierarchical neural coding scheme will roughly require $N=b \, \log_b R$ neurons.   To derive the optimal base (i.e. the base that minimizes the number of  the neurons), we evaluate the extremum $\partial N / \partial b = 0$:
\begin{equation}
\partial N / \partial b= \frac{\partial (b \, \log_b R)}{ \partial b}=\frac{ \partial (\frac{b \, \ln R}{ln b})} {\partial b} = \ln R \, \,  \frac{\ln b -1 }{(\ln b)^2}
\end{equation}
Setting  $\partial N / \partial b = 0$ gives  $\ln b - 1=0$. Therefore the number of neurons is extremized when  $b=e$.  It is easy to check that this is a minimum.

\section{Optimizing the grid system: winner-take-all decoder} 
\label{s:wta}
\subsection{Lagrange multiplier approach}
\label{s:lagrange}
We saw in the main text that, for a winner-take-all decoder, the problem of deriving the optimal ratios of adjacent grid scales in one dimension is equivalent to minimizing the sum of a set of numbers ($N =  d \sum_{i=1}^m r_i$)  while fixing the product  ($R_1 = \prod_{i=1}^m r_i$) to take the fixed value $R$.   Mathematically, it is equivalent to minimize $N$ while fixing  $\ln R$.     When $N$ is large we can treat it as a continuous variable and use the method of Lagrange multipliers as follows.
First,we  construct the auxiliary function $H(r_1 \cdots r_N, \beta) = N - \beta \, (\ln R_1 - \ln R)$
and then extremize $H$ with respect to each $r_i$ and $\beta$.    Extremizing with respect to $r_i$ gives
\begin{equation}
{\partial H \over  \partial {r_i}}= d-\frac{\beta}{r_i}=0  ~~~~~\Longrightarrow~~~~~
r_i = {\beta  \over d} \equiv r \, .
\end{equation}
Next, extremizing with respect to $\beta$ to implement the constraint on the resolution gives
\begin{equation}
{\partial H \over \partial \beta} 
=\ln R_1-\ln R=m \ln r - \ln R =0 ~~~~~\Longrightarrow~~~~~   r =  R^{1/m}
\end{equation}
Having thus implemented the constraint that $\ln R_1 = \ln R$ , it follows that $H = N = d \, m \, R^{1/m}$.    Alternatively, solving for $m$ in terms of $r$, we can write $H =  d \ r \, (\ln R )\, / \ln r) = d \, r  \log_r R$.  It remains to minimize the number of cells $N$ with respect to $r$,
% i.e. $\partial \ln N / \partial m = 0$. Equavelently,
\begin{equation}
{\partial H \over \partial r} =  d \, \ln  R  \left[ {1 \over \ln r} - \left(1 \over \ln r \right)^2 \right] = 0 
~~~~~\Longrightarrow~~~~~  \ln r = 1
\end{equation}
This is in turn implies our result
\begin{equation}
r = e
\end{equation}
for the optimal ratio between adjacent scales in a hierarchical, grid coding scheme for position in one dimension, using a winner-take-all decoder.
%$$
%\partial \ln N / \partial m=(d/\alpha) \partial (\ln m+\frac{1}{m} \ln S_0) / \partial m= (d/\alpha) (\frac{1}{m}-\frac{1}{m^2} \ln S_0)=0
%$$
%This gives $m = \ln S_0$.   But since the resolution is $S_0 = r^m$, it follows that the optimal ratio must be $r=e$! 
%Therefore, the optimal scaling factor is $e$.
In this argument we employed the sleight of hand that $N$ and $m$ can be treated as continuous variables, which is approximately valid when $N$ is large.  This condition obtains if the required resolution $R$ is large.   A more careful argument is given below that preserves the integer character of $N$ and $m$.

\subsection{Integer $N$ and $m$}
\label{s:integer}
As discussed above, we seek to minimize the sum of a set of numbers ($N =  d \sum_{i=1}^m r_i$)  while fixing the product  ($R =  \prod_{i=1}^m r_i$) to take a fixed value.    We wish to carry out this minimization while recognizing that the number of neurons is an integer.  First, consider the arithmetic mean-geometric mean inequality which states that, for a set of non-negative real numbers, ${x_1,x_2,...,x_m}$, the following holds:
\begin{equation}
(x_1+x_2+...+x_m)/m\geq (x_1x_2...x_m)^{1/m} \, ,
\end{equation}
with equality if and only if all the $x_i$'s are equal.  Applying this inequality, it is easy to see that to minimize $\sum_{i=1}^{m} r_i$, all of the $r_i$ should be equal. We denote this common value as $r$, and we can write $r= R^{1/m}$.

Therefore, we have
\begin{equation}
N= d \sum_{i=1}^{m} r= m \, d  \, R^{1/m}  
\end{equation}
Suppose $ R=e^{z+\epsilon}$, where $z$ is an integer, and $\epsilon \in[0,1)$. By taking the first derivative of N with respect to m, and setting it to zero, we find that $N$ is minimized when $m=z+\epsilon$.    However, since $m$ is an integer the minimum will be achieved either at $m=z$ or $m=z+1$. (Here we used the fact  $m  R^{1/m}$ is monotonically increasing between $0$ and $z+\epsilon$ and is monotonically decreasing between $z + \epsilon$ and $\infty$.)
Thus, minimizing N requires either
\begin{equation}
r=(e^{z+\epsilon})^{\frac{1}{z}}=e^{\frac{z+\epsilon}{z}} 
~~~~~
{\rm or} 
~~~~~ 
r=(e^{z+\epsilon})^{\frac{1}{z+1}}=e^{\frac{z+\epsilon}{z+1}} \, .
\end{equation}
In either case, when $z$ is large (and therefore $R$, $N$ and $m$ are large), $r \rightarrow e$.  This shows that when the resolution $R$ is sufficiently large,  the total number of neurons $N$ is minimized when $r_i \approx e$ for all $i$. 
\section{Optimizing the grid system: Bayesian decoder}
\subsection{Neuron number and resolution}
\label{s:bayesian}
In the main text we argued that the optimal scale factor in one dimension is $r = e$ assuming that decoding is based on the responses of the most active cell at each scale. However, the decoding strategy could use more information from the population of neurons.  Thus, we consider a Bayes-optimal decoder that accounts for \emph{all} available information by forming a posterior distribution of position, given the activity of all grid cells in the population. We can make quantitative predictions in this general setting if we assume that the firing of different grid cells is statistically independent and that the tuning curves at each scale $i$ provide dense, uniform, coverage of the interval $\lambda_i$. With these assumptions, the posterior distribution of the animal's position, given the activity of grid cells at the single scale $i$, $P(x \, | \,i)$, may be approximated by a series of Gaussian bumps of standard deviation $\sigma_i$ spaced at the period $\lambda_i$. Furthermore, $\sigma_i = c d^{-1/2} l_i$, where $l_i$ is the width of each tuning curve, $c$ is a dimensionless factor incorporating the tuning curve shape and  noisiness of single neurons, and $d$ is the coverage factor.  The linear dependence on $l_i$ follows from dimensional analysis. From the definition of $d$ given in the main text, $d = n_i \frac{l_i}{\lambda_i}$, we see that $d$ can be interpreted as the number of cells with tuning curves overlapping a given point in space. The square-root dependence of $\sigma_i$ on $d$ then follows, as this is the effective number of neurons independently encoding position. We assume here that $d$ is large; this is necessary for the Gaussian approximation to hold. Finally, combining the equation for $\sigma$ with the relationship, $n_{i} = d \frac{\lambda_{i}}{l_{i}}$, gives $n_{i} = c\sqrt{d} \frac{\lambda_{i}}{\sigma_{i}}$. Therefore, the total number of neurons, which we would like to minimize, is $N = c\sqrt{d} \sum_{i=1}^m \frac{\lambda_i}{\sigma_i}$.

In the main text, we minimized $N$ while fixing the resolution $R_1$. In our present Bayesian decoding model, $R_1$ will be related to the standard deviation $\delta_m$ of the distribution of location $x$ given the activity of all $m$ scales, $Q_m(x)$.  In general, the activity of the grid cells at all scales larger than $\lambda_{i}$ provides a distribution over position $Q_{i-1}(x)$ which is combined with the posterior $P(x \, | \,i)$ to find the distribution $Q_i(x)$ given all scales $1$ to $i$. Since we assume independence across scales, $Q_{i-1}(x)$ is obtained by taking the product over all the posteriors up to scale $i-1$: $Q_{i-1}(x) = {\cal N} \prod_{j=1}^{i-1} P(x\,|\,j)$, where ${\cal N}$ normalizes the distribution. Furthermore,  $Q_{i}(x) = {\cal N}' \, P(x\,|\,i) \, Q_{i-1}(x)$. The posteriors from different scales have different periodicities, so multiplying them against each other will tend to suppress all peaks except the central one, which is aligned across scales. We may thus approximate $Q_{i-1}(x)$ and $Q_i(x)$ by single Gaussians whose standard deviations we will denote as $\delta_{i-1}$ and $\delta_{i}$, respectively. The validity of this approximation is taken up in further detail in section \ref{s:scalefactor} below.  By dimensional analysis, $\delta_{i} = \delta_{i-1} / \rho(\frac{\lambda_{i}}{\sigma_{i}}, \frac{\sigma_{i}}{\delta_{i-1}})$. With the stated Gaussianity assumptions, the function $\rho$ may be explicitly defined and evaluated numerically (section \ref{s:scalefactor}). A Bayes-optimal decoder will then estimate the animal's position with error proportional to the posterior standard deviation over all $m$ scales, $\delta_{m} = \left(\prod_{i} \rho_{i}\right)^{-1} \sigma_{1}$, and no unbiased decoder can do better than this. (We are abbreviating $\rho_i \equiv \rho(\lambda_i/\sigma_i, \sigma_i/\delta_{i-1}$.)   Thus, the resolution constraint imposed in the main text becomes, in the present context, a constraint on $\prod_i \rho_i$. We will show below that $\rho$ is in fact equal to the scale factor $r_i = \lambda_i / \lambda_{i+1}$.

\begin{figure}
\centering
\includegraphics[keepaspectratio,width=0.4\linewidth]{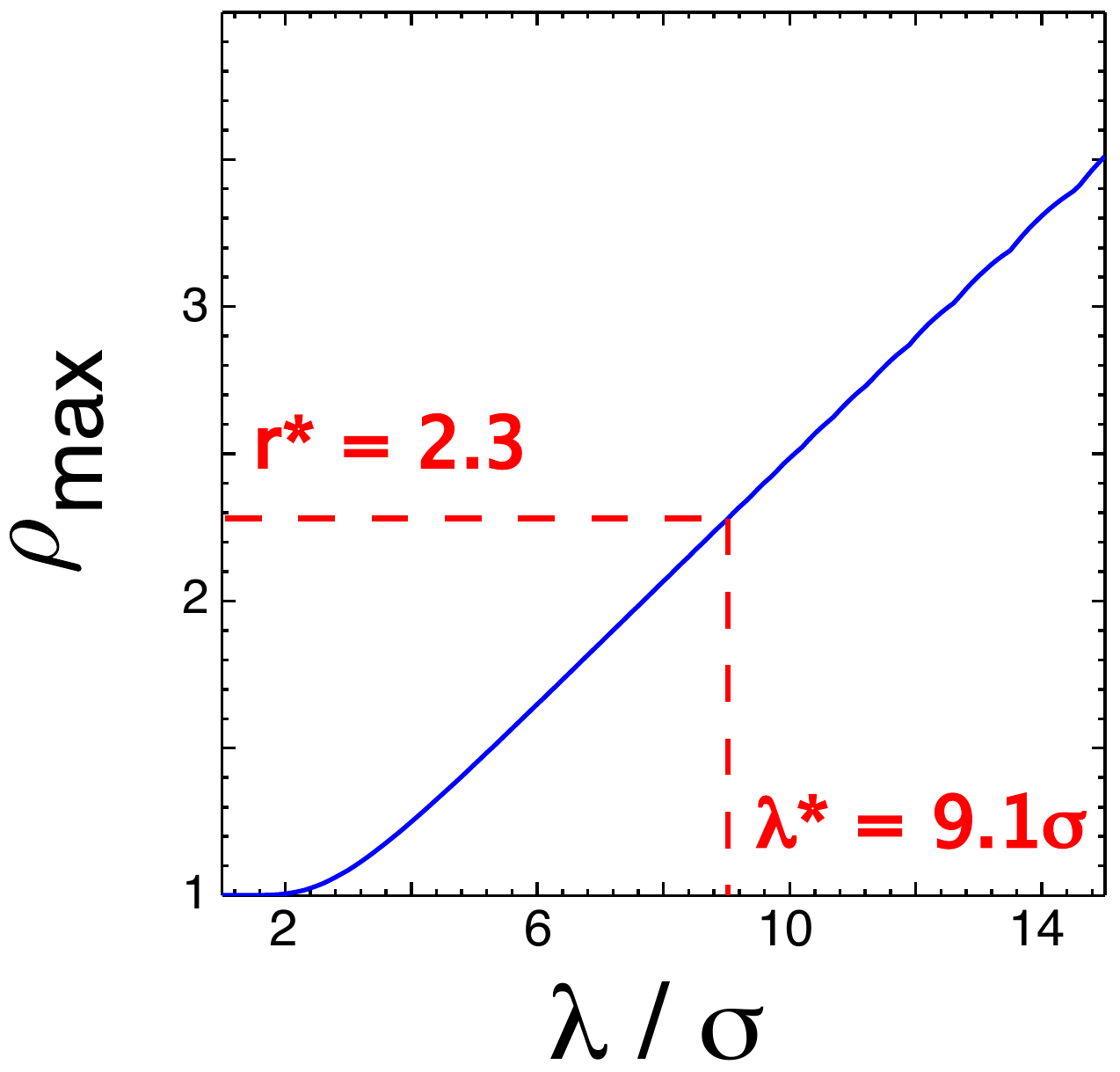}
%\hspace{0.03\linewidth}
\caption{
{$\rho_{max} \equiv {\max_{\sigma / \delta}} \, \rho(\frac{\lambda}{\sigma}, \frac{\sigma}{\delta})$ is the scaling factor after optimizing $N$ over $\sigma / \delta$. The values $r^*$ and $\lambda^*$ are the values chosen by the complete optimization procedure.
}}
\label{fig:rmax}
\end{figure}

Thus, we would like to minimize $N = c\sqrt{d} \sum_{i=1}^m \frac{\lambda_i}{\sigma_i}$ subject to a constraint $R = \prod_{i} \rho(\frac{\lambda_{i}}{\sigma_{i}}, \frac{\sigma_{i}}{\delta_{i-1}})$. The minimization is with respect to the parameters $\lambda_i/\sigma_i$ and $\sigma_i/\delta_{i-1}$.
We perform the calculation in two steps: first optimizing over 
$\sigma_i/\delta_{i-1}$, then over $\lambda_i/\sigma_i$.
The former parameter only affects $N$ indirectly, by changing the number of scales $m$ through the constraint $\prod_{i=1}^m \rho(\frac{\lambda_{i}}{\sigma_{i}}, \frac{\sigma_{i}}{\delta_{i-1}})$. Choosing 
$\sigma_i/\delta_{i-1}$
 to maximize $\rho$ will minimize $m$, and therefore $N$. We thus replace $\rho$ by 
  $\rho_{max}(\lambda / \sigma) \equiv {\max_{\sigma / \delta}} \, \rho(\lambda/\sigma,\sigma/\delta)$
% $\rho_{max}(\lambda / \sigma) \equiv {\max_{\sigma / \delta}} \, \rho(\frac{\lambda}{\sigma}, \frac{\sigma}{\delta})$ 
 and minimize $N$ over the remaining parameters  
$\lambda_i/\sigma_i$.
As in the main text, the problem has a symmetry under permutations of the $i$, so the optimal $\lambda_{i} / \sigma_{i}$ and $\sigma_i / \delta_{i-1}$ are independent of $i$. Thus, $m = \ln R / \ln \rho_{max}$ and $N \propto \frac{\lambda / \sigma}{\ln \rho_{max}(\lambda / \sigma)}$. We can invert the one-to-one relationship between $\rho_{max}$ and $\lambda / \sigma$ (Fig. \ref{fig:rmax}), and minimize $N$ over $\rho_{max}$ to get $\rho_{max}^* = 2.3$. In fact, $\rho$ is equal to the scale factor: $\rho_i = r_i = \lambda_i / \lambda_{i+1}$. To see this, express $\rho_i$ as a product: 
%$\rho_i = \delta_{i-1}/\delta_i = (\delta_{i-1} / \sigma_i) (\sigma_i / \lambda_i) (\lambda_i / \lambda_{i+1}) (\lambda_{i+1} / \sigma_{i+1}) (\sigma_{i+1} / \delta_i)$
$\rho_i = \frac{\delta_{i-1}}{\delta_{i}} = \frac{\delta_{i-1}}{\sigma_{i}}\frac{\sigma_{i}}{\lambda_{i}}\frac{\lambda_{i}}{\lambda_{i+1}}\frac{\lambda_{i+1}}{\sigma_{i+1}}\frac{\sigma_{i+1}}{\delta_i} $.
 Since the factors $\sigma_i / \delta_{i-1}$ and $\lambda_i / \sigma_i$ are independent of $i$, they cancel in the product and we are left with $\rho_i = \lambda_i/\lambda_{i+1}$.
%$\rho_i = \frac{\lambda_i}{\lambda_{i+1}}$.

We have thus seen that the Bayesian decoder predicts an optimal scaling factor $r^* = 2.3$ in one dimension. This is similar to, but somewhat different than, the winner-take-all result $r^* = e = 2.7$.  At a technical level the difference arises from the fact that the function $\rho_{max}(\lambda / \sigma)$ does not satisfy $\rho_{max} = \frac{\lambda}{\sigma}$ as used previously, but is instead more nearly approximated by a linear function with an offset: $\rho \approx \alpha^{-1} (\frac{\lambda}{\sigma} + \beta)$.   A more conceptual reason for the difference is that the Gaussian posterior used here has long tails which are absent in the case with compact firing fields. The scale factor must then be smaller to keep the ambiguous secondary peaks of the next scale far enough into the tails to be adequately suppressed.   The optimization also predicts $\lambda^* = 9.1 \, \sigma$, which may be combined with the formula $\sigma = c d^{-1/2} l$ to predict $l/\lambda$. However, this relationship depends on the parameters $c$ and $d$ which may only be calculated from a more detailed description of the single neuron response properties.  For this reason, the general Bayesian analysis above does not predict the ratio of the grid periodicity to the width of individual grid fields.  Note that $\lambda^* = 9.1 \, \sigma$ also implies  that $\sigma_i / \lambda_{i+1} \approx 4$ -- i.e. that the peaks of the  posterior distribution at scale $i+1$ are separated by ~4 of the standard deviations of the peaks at scale $i$.

A similar Bayesian analysis can be carried out for two dimensional grid fields. The posteriors $P(x\,|\,i)$ become two-dimensional sums-of-Gaussians, with the centers of the Gaussians laid out on the vertices of the grid. $Q_i(x)$ is then similarly approximated by a two-dimensional Gaussian. The form of the function $\rho$ changes (section \ref{s:scalefactor}), but the logic of the above derivation is otherwise unaltered.

\subsection{Calculating $\rho(\frac{\lambda}{\sigma}, \frac{\sigma}{\delta})$}
\label{s:scalefactor}
Section \ref{s:bayesian} argued that  the function $\rho(\frac{\lambda}{\sigma},\frac{\sigma}{\delta})$ can be computed by making the approximation that the posterior distribution of the animal's position given the activity at a single scale $i$, $P(x\,|\,i)$, is a periodic sum-of-Gaussians:
\begin{equation}
P(x\, | \,i) = {1 \over 2K + 1} \, \sum_{n=-K}^{K} \frac{1}{\sqrt{2\pi\sigma_{i}^2}} e^{-\frac{1}{2\sigma_{i}^2} \left(x - n\lambda_i\right)^2} \, 
\end{equation}
where $K$ is assumed is large.
We further approximate the posterior given the activity of \emph{all} scales coarser than $\lambda_{i}$ by a Gaussian with standard deviation $\delta_{i-1}$:
\begin{equation}
Q_{i-1}(x) = \frac{1}{\sqrt{2\pi\delta_{i-1}^2}} e^{-x^2 / 2\delta_{i-1}^2}
\end{equation}
Assuming independence across scales, it then follows that $Q_{i}(x) = \frac{P(x\,|\,i)Q_{i-1}(x)}{\int \, \mathrm{d}x P(x\,|\,i)Q_{i-1}(x)}$. Then $\rho(\lambda_i/\sigma_i, \sigma_i / \delta_{i-1})$ is given by $\delta_{i-1} / \delta_{i}$, where $\delta_{i}$ is the standard deviation of $Q_{i}$. We therefore must calculate $Q_{i}(x)$ and its variance in order to obtain $\rho$. After some algebraic manipulation, we find,
\begin{equation}
Q_{i}(x) = \sum_{n=-K}^{K}  \pi_n \frac{1}{\sqrt{2\pi\Sigma^2}} e^{-(x - \mu_n)^2 / 2\Sigma^2},
\end{equation}
where $\Sigma^2 = \left(\sigma_i^{-2} + \delta_{i-1}^{-2}\right)^{-1}$, $\mu_n = \left(\frac{\Sigma}{\sigma_i}\right)^2 \lambda_i \, n$, and
\begin{equation}
\pi_n = \frac{1}{Z} e^{-n^2 \lambda_i^2 / 2(\sigma_i^2 + \delta_{i-1}^2)}.
\end{equation}
$Z$ is a normalization factor enforcing $\sum_n \pi_n = 1$. $Q_{i}$ is thus a mixture-of-Gaussians, seemingly contradicting our approximation that all the $Q$ are Gaussian. However, if the secondary peaks of $P(x\,|\,i)$ are well into the tails of $Q_{i-1}(x)$, then they will be suppressed (quantitatively, if $\lambda_i^2 \gg \sigma_i^2 + \delta_{i-1}^2$, then $\pi_n \ll \pi_0$ for $|n| \ge 1$), so that our assumed Gaussian form for $Q$ holds to a good approximation. In particular, at the values of $\lambda, \sigma,$ and $\delta$ selected by the optimization procedure described in section \ref{s:bayesian}, $\pi_1 = 1.3\cdot10^{-3}\pi_0$. So our approximation is self-consistent.

Next, we find the variance $\delta_{i}^2$:
\begin{align}
\delta_{i}^2 &= \langle x^2 \rangle_{Q_{i}}  \nonumber \\
&= \sum_n \pi_n (\Sigma^2 + \mu_n^2) \nonumber  \\
&= \Sigma^2 \left(1 + \left(\frac{\Sigma}{\sigma_i}\right)^2\left(\frac{\lambda_i}{\sigma_i}\right)^2 \sum_n n^2 \pi_n\right) \nonumber \\
&= \delta_{i-1}^2 \left(1 + \frac{\delta_{i-1}^2}{\sigma_i^2}\right)^{-1} \left(1 + \left(\frac{\Sigma}{\sigma_i}\right)^2\left(\frac{\lambda_i}{\sigma_i}\right)^2 \sum_n n^2 \pi_n\right).
\end{align}
We can finally read off $\rho(\frac{\lambda_i}{\sigma_i}, \frac{\sigma_i}{\delta_{i-1}})$ as the ratio $\delta_{i-1} / \delta_{i}$:
\begin{equation}
\rho(\frac{\lambda_i}{\sigma_i}, \frac{\sigma_i}{\delta_{i-1}}) = \left(1 + \frac{\delta_{i-1}^2}{\sigma_i^2}\right)^{1/2} \left(1 + \left(1 + \frac{\sigma_i^2}{\delta_{i-1}^2}\right)^{-1}\left(\frac{\lambda_i}{\sigma_i}\right)^2 \sum_n n^2 \pi_n\right)^{-1/2}.
\end{equation}
 For the calculations reported in the text, we took $K = 500$. 

Section \ref{s:bayesian} explained that we are interested in maximizing $\rho$ over $\frac{\sigma}{\delta}$, holding $\frac{\lambda}{\sigma}$ fixed. The first factor in $\rho$ increases monotonically with decreasing $\frac{\sigma}{\delta}$; however, $\sum_n n^2 \pi_n$ also increases and this has the effect of reducing $\rho$. The optimal $\frac{\sigma}{\delta}$ is thus controlled by a tradeoff between these factors. The first factor is related to the increasing precision given by narrowing the central peak of $P(x\,|\,i)$, while the second factor describes the ambiguity from multiple peaks.
 
 The derivation can be repeated in the two-dimensional case. We take $P(x\,|\,i)$ to be a sum-of-Gaussians with peaks centered on the vertices of a regular lattice generated by the vectors $(\lambda_i \hat{u}, \lambda_i \vec{v})$. We also define $\delta_i^2 \equiv \frac{1}{2} \langle|x|^2\rangle_{Q_i}$. The factor of $1/2$ ensures that the variance so defined is measured as an average over the two dimensions of space.  The derivation is otherwise parallel to the above, and the result is,
\begin{equation}
\rho_2(\frac{\lambda_i}{\sigma_i}, \frac{\sigma_i}{\delta_{i-1}}) = \left(1 + \frac{\delta_{i-1}^2}{\sigma_i^2}\right)^{1/2} \left(2 + \left(1 + \frac{\sigma_i^2}{\delta_{i-1}^2}\right)^{-1}\left(\frac{\lambda_i}{\sigma_i}\right)^2 \sum_{n,m} |n\hat{u} + m\vec{v}|^2\,\,\pi_{n,m}\right)^{-1/2},
\end{equation}
where $\pi_{n,m} = \frac{1}{Z} e^{-|n\hat{u} + m\vec{v}|^2  \lambda_i^2 / 2(\sigma_i^2 + \delta_{i-1}^2)}$.

 \section{Reanalysis of grid  data from previous studies}
\label{s:reanalysis}
We reanalyzed the data from Barry et. al \cite{Barry2007} and Stensola et. al\cite{Stensola2012} in order to get the mean and the variance of the ratio of adjacent grid scales.   For Barry et. al\cite{Barry2007}, we first read the raw data from Figure 3b of the main text using the software GraphClick, which  allows  retrieval of the original (x,y)-coordinates from the image.   This gave the scales of grid cells recorded from 6 different rats.   For each animal, we grouped the grids that had similar periodicities (i.e. differed by less than 20\%) and calculated the mean periodicity for each group. We defined this mean periodicity as the scale of each group.  For 4 out of 6 rats, there were 2 scales in the data. For 1 out  6 rats, there were 3 grid scales. For the remaining rat, only 1 scale was obtained as only 1 cell was recorded from that rat. We excluded this rat from further analysis. We then calculated the ratio between adjacent grid scales, resulting in 6 ratios from 5 rats. The mean and variance of the ratio were $1.64$ and $0.09$, respectively ($n = 6$). 

For Stensola et. al\cite{Stensola2012},  we first read in the data using GraphClick from Figure 5d of the main text.  This gave the scale ratios between different grids for 16 different rats. We then pooled all the ratios together and calculated the mean and variance.  The mean and variance of the ratio were $1.42$ and $0.17$, respectively ($n = 24$). 

Giocomo et. al\cite{Giocomo2011} reported the ratios between the grid period and the \emph{radius} of grid field (measured as the radius of the circle around the center field of the autocorrelation map of the grid cells ) to be $3.26 \pm 0.07$ and $3.32 \pm 0.06$ for Wild-type and HCN KO mice, respecitvely. We linearly transform these measurements to the ratios between grid period and the \emph{diameter} of the grid field to facilitate the comparison to our theoretical predictions. The results are plotted in a bar graph (Fig. 4B in the main text).

Finally, in Figure 4C, we replotted Fig.~1c from \cite{Hafting2005} by  reading in the data using GraphClick and then translating that information back into a plot.

\section{General optimality of the triangular lattice}
\label{s:triangular}
Our task is to minimize the number of neurons in a population made up of $m$ modules, $N = d \sum_{i=1}^m |v_\perp| (\frac{\lambda_i}{l_i})^2$, subject to a constraint on resolution $R = F(\{\lambda, l, \mathbf{u}, \mathbf{v}\}, m)$. The specific form of the resolution function $F$ will, of course, depend on the details of tuning curve shape, noise, and decoder performance. Nevertheless, we will prove that the triangular lattice is optimal in \emph{all} models sharing the following general properties:
\begin{itemize}
\item \textbf{Uniqueness:} Our optimization problem has a unique solution for all $R$. The optimal parameters are continuous functions of $R$. 

\item \textbf{Symmetry:} Simultaneous \emph{rotation} of all firing rate maps leaves $F$ invariant. Likewise, $F$ is invariant under simultaneous \emph{rescaling} of all maps. These transformations are manifestly symmetries of the neuron number $N$. Rotation invariance implies that $F$ depends on $\mathbf{u}$ and $\mathbf{v}$ only through the two scalar parameters $v_\perp$ and $v_\parallel$ (the components of $\mathbf{v}$ orthogonal to and parallel to $\mathbf{u}$, respectively). Scale invariance implies that the dependence on the dimensionful parameters $\{\lambda, l\}$ is only through the ratios $\{r, \lambda / l\}$, where $r_i = \lambda_{i} / \lambda_{i+1}$ are the scale factors. The resolution formulas in both the winner-take-all and the Bayesian formulations are evidently scale-invariant, as they depend only on dimensionless ratios of grid parameters. We will also assume that firing fields are circularly-symmetric.

\item \textbf{Asymptotics:} The resolution $F(\{r, \lambda / l\}, v_\parallel, v_\perp, m)$ increases monotonically with each $\lambda_i / l_i$. When all $\lambda_i / l_i \to \infty$, the grid cells are effectively place cells and so the grid geometry cannot matter. Therefore, $F$ becomes independent of $\mathbf{v}$ in this limit.

\end{itemize}

We will first argue that the uniqueness and symmetry properties imply that the optimal lattice can only be square or triangular. The asymptotic condition then picks out the triangular grid as the better of these two. To see the implications of the symmetry condition, consider the following transformation of the parameters:
\begin{align*}
v_\perp &\to -v_\perp / |\mathbf{v}|^2 \\
v_\parallel &\to v_\parallel / |\mathbf{v}|^2 \\
l_i &\to l_i / |\mathbf{v}| 
\end{align*}
This takes the vector $\mathbf{v}$, reflects it through $\mathbf{u}$ (keeping the same angle with $\mathbf{u}$), and scales it to have length $1 / |\mathbf{v}|$. This new $\mathbf{v}$, together with $\mathbf{u}$, thus generates the same lattice as the original $\mathbf{u}$ and $\mathbf{v}$, but rotated, scaled, and with the roles of $\mathbf{u}$ and $\mathbf{v}$ exchanged. We then also scale all field width parameters by the same factor $1 / |\mathbf{v}|$ to compensate for the stretching of the lattice. And although this is a rotation of the lattice and not the firing fields, our assumed isotropy of the firing fields implies that the transformation is indistinguishable from a rotation of the entire rate map. Since the overall transformation is equivalent to a common rotation and scaling of all rate maps, it will (by our symmetry assumption) leave the neuron number and resolution unchanged. If the optimal lattice is unique, it must then be invariant under this transformation.

Which lattices are invariant under the above transformation? It must take the generator $\mathbf{v}$ to another generator $\mathbf{v'}$ of the \emph{same} lattice. This requirement demands that the generators are related by a modular transformation:
\begin{align*}
\mathbf{v'} &= a \mathbf{v} + b \mathbf{u} \\
\mathbf{u} &= c \mathbf{v} + d \mathbf{u},
\end{align*}
with $a, b, c, d$ integers such that $|ad - bc| = 1$. The second equation, and linear independence of $\mathbf{u}$ and $\mathbf{v}$, require $c = 0, d = 1$ and so $|a| = 1$. Plugging in our transformation of $\mathbf{v}$, the first equation then gives $a=-1$, $|\mathbf{v}| = 1$ and $v_\parallel = b / 2$. Since $\mathbf{v} + n\mathbf{u}$ will generate the same lattice as $\mathbf{v}$, for any integer $n$, we may assume $0 \le v_\parallel < 1$. The only solutions are the square lattice with $v_\parallel = 0, v_\perp = 1$ and the triangular lattice with $v_\parallel = 1/2, v_\perp = \sqrt{3}/2$.

It remains to choose between these two possibilities. We want to minimize $N = d \sum_i |v_\perp| (\frac{\lambda_i}{l_i})^2$, so it seems that we should minimize $|v_\perp|$, giving the triangular lattice. However, the constraint on resolution will introduce $\mathbf{v}-$dependence into $\lambda / l$, so it is not immediately clear that we can minimize $N$ by minimizing $|v_\perp|$ alone. But the asymptotic condition implies the existence of a large-$R$ regime tied to large $\lambda / l$, and asserts that in this limit the $\mathbf{v}$-dependence drops out. Therefore,  the triangular lattice is optimal for large enough $R$. Since the only other possible optimum is the square lattice, and our uniqueness assumption prevents the solution from changing discontinuously as $R$ is lowered, it must be the case that the triangular lattice is optimal for \emph{all} $R$.

\end{document}